\documentclass{article}

\pdfoutput=1

\usepackage{authblk}
\usepackage[a4paper, total={5in, 8in}]{geometry}
\usepackage{graphicx} 
\title{A Binary Annular Phase Mask to Regulate Spherical Aberration and Allow Super-Localization in Single-Particle Tracking over Extended Depth-of-Focus}

\author[a,b]{Quentin Gresil}

\author[a,b]{Antony Lee \footnote{Present adress: Laboratoire Physico-Chimie Curie, Institut Curie, CNRS - UMR 168, Sorbonne Université 75005, Paris, France}}

\author[c]{Olivier L\'ev\^eque}

\author[a,b]{Karen Caicedo}

\author[a,b]{Blanca Martín Muñoz}

\author[c]{Caroline Kulcs\'ar}

\author[c]{François Goudail}

\author[d]{Pierre Bon}

\author[a,b]{Laurent Cognet \footnote{Corresponding author, laurent.cognet@u-bordeaux.fr}}

\affil[a]{\small \textit{Laboratoire Photonique Numerique et Nanoscience, Universite de Bordeaux, 33400, Talence, France}}

\affil[b]{\small \textit{LP2N, Institut d’Optique Graduate School, CNRS - UMR 5298, 33400, Talence, France}}

\affil[c]{\small \textit{Université Paris-Saclay, Institut d’Optique Graduate School, CNRS, Laboratoire Charles Fabry, 91127, Palaiseau, France}}

\affil[d]{\small \textit{XLIM, CNRS - UMR 7252, Université de Limoges, 87000, Limoges, France}}

\date{}

\begin{document}

\maketitle

\begin{abstract}
Important applications of single-particle tracking (SPT) aim at deciphering the diffusion properties of single fluorescent  nanoparticles immersed in heterogeneous environments, such as multi-cellular biological tissues. 
To maximize the particle localization precision in such complex environments,  high numerical aperture objectives are often required,  which intrinsically restrict depth-of-focus (DOF) to less than a micrometer and impedes recording long trajectories when particles escape the plane of focus. 
In this work, we show that a simple binary phase mask can work with the spherical aberration inevitably induced  by thick sample inhomogeneities, to extend the DOF of a single-molecule fluorescence microscope over more than 4 \textmu m. The effect of point-spread-function (PSF) engineering over spherical aberration regularizes inhomogeneities of the PSF along the optical axis by restricting it to a narrow distribution. This allows the use of a single fitting function (i.e. Gaussian function) to localize single emitters over the whole extended DOF.
Application of this simple approach on diffusing  nanoparticles demonstrate that SPT trajectories  can be recorded on significantly longer times.\\
 
\end{abstract}

\section{Introduction}
\label{introduction}

Single-particle tracking (SPT) and localization-based super-resolution microscopy share the same advantage of allowing single emitters to be localized with sub-diffraction precision - down to nanometer scales -  after fitting  emitter detected signals by the point spread function (PSF) of the microscope \cite{godin_super-resolution_2014}. In both cases, imaging single emitters is performed under the constraint of low photon budgets, which limits the signal to noise ratio at which the images are formed, and affects the precision of emitter super-localization \cite{mortensen_optimized_2010}. In many instances, microscope PSFs are approximated by Gaussian functions allowing efficient localizations to be obtained \cite{lee_unraveling_2017}. In these applications, high NA objectives are typically used to collect the maximum number of photons but they inherently limit the depth-of-focus (DOF) of the imaging instrument. This limited DOF is however particularly detrimental for SPT applications, especially when aiming at studying 3D diffusion processes in deep complex samples, where detected emitters frequently escape the imaging DOF. In this context, several approaches have been proposed to extend the DOF, either based on multi-plane imaging \cite{ram_high_2008} or on PSF engineering coupled with complex PSF fitting \cite{pavani_three-dimensional_2009,shechtman_optimal_2014}. In general, while such approaches also aim at performing super-localization in 3D, they require quite large photon budgets for being effective. In the context of conventional widefield diffraction-limited fluorescent microscopy, extended DOF (eDOF) has been proposed \cite{dowski_extended_1995, abrahamsson_new_2006}, including  when sample index mismatches induce spherical aberrations \cite{king_performance_2018}. Interestingly, the case of super-resolution and SPT performed with eDOF deep into a sample -i.e. on large distance far from the surface of a thick sample - has not been directly addressed in the presence of spherical aberrations, which by definition can neither be ignored nor corrected uniformly on large arbitrary axial distances. To address this point, we propose here a simple approach based on the introduction of binary annular phase masks \cite{leveque_co-designed_2020, leveque_validity_2022} to extend the DOF for SPT applications in the realistic case of spherical aberration and 
using a simple fitting function.
More precisely, we show that the mask regularizes PSF spherical aberration over eDOF which allows super-localization with commonly used 2D Gaussian fitting which does not depend on the distance of the molecule from the imaging plane. We demonstrate the effectiveness of the approach on SPT data where recorded trajectory lengths are increased in a single experiment.


\section{Results}
\label{results}

\subsection{DOF of single emitter superlocalization in the presence of spherical aberrations}
\label{section_sa}

When an emitter is inserted in a thick sample and imaged through this sample by high NA objectives, spherical aberrations generally arise due to the index mismatch introduced by the sample and vary depending on the axial position of the emitter \cite{Gibson:91}. It is thus challenging to compensate spherical aberrations simultaneously at different depths. For instance, this precludes to image several emitters in the volume of samples displaying unknown local optical inhomogeneities like biological tissue do. Note that real-time adaptive optics was developed to compensate such aberrations, but only at one imaging plane at a time, and yet with complex instrumentation to analyse the aberration present at a given focal plane \cite{booth_adaptive_2014}. 

\begin{figure}
    \centering
    \includegraphics[scale=1]{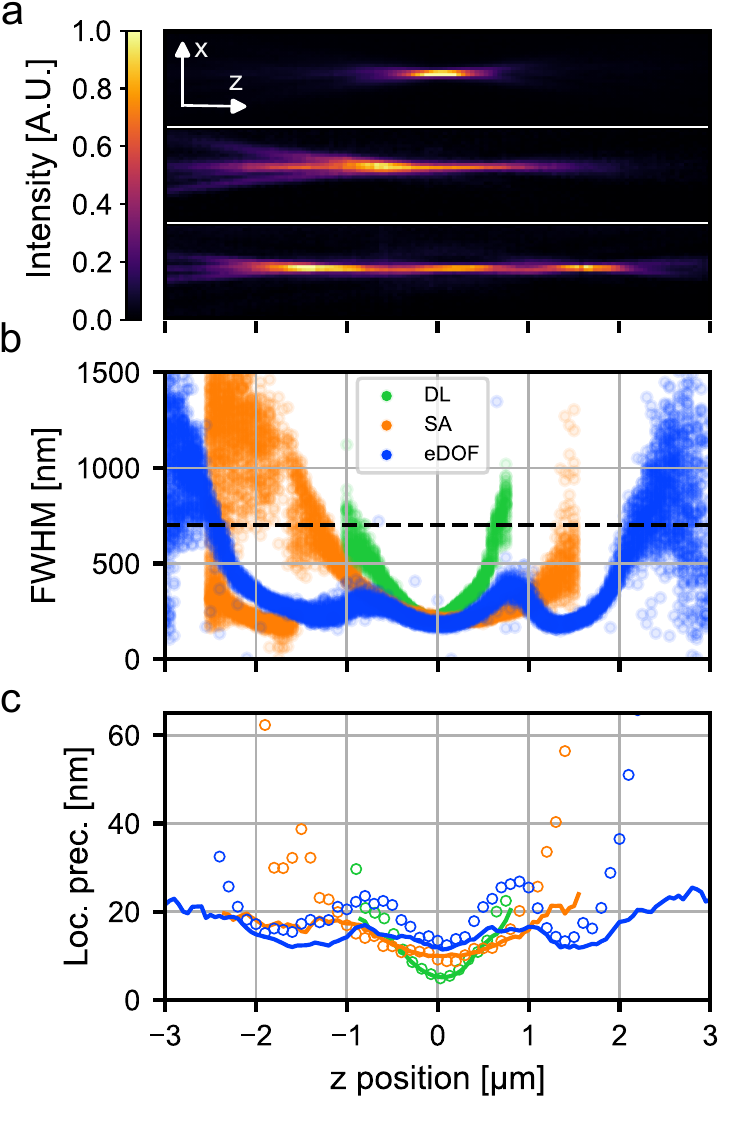}
    \caption{\textbf{Depth-of-focus and localization precision of diffraction-limited (DL), spherical-aberrated (SA) and extended depth-of-focus (eDOF) PSFs.} \textbf{(a)} Measured axial intensity profiles of DL (top panel), SA (middle panel) and eDOF in the presence of spherical aberrations (bottom panel) PSFs. 1 \textmu m scale bar. \textbf{(b)} FWHM of Gaussian fits along the depth-of-focus of the three recorded PSFs. The dotted horizontal black line accounts for the threshold applied in (c). \textbf{(c)} Computed localization precision obtained from Gaussian fits along the depth-of-focus. Full lines dipslay the Cramer-Rao Lower Bound (CRLB) computed for the three different PSFs.}
    \label{locpres}
\end{figure}

We thus aimed at investigating the impact of spherical aberration on the super-localization of single particles with the additional constraint that super-localization relies on the use of a unique and simple PSF fitting function along the DOF. We chose 2D Gaussian fitting which is commonly used in the realistic situation of a limited photon number budget such as in SPT experiments performed in thick biological samples \cite{godin_single-nanotube_2017}. We therefore rejected the use of complex fitting procedures to adapt uncontrolled aberrated PSFs generated by the sample. This choice was also motivated by the performance and broad code availability of Gaussian fitting approaches to super-localize single emitters \cite{ovesny_thunderstorm_2014}.

For this purpose, a series of images of single fluorescent nanoparticles in water immobilized on a glass coverslip was first acquired as a function of defocus in the absence or presence of spherical aberrations (but without phase mask). More precisely, we acquired z-stacks of single fluorescent particle at a photon budget of 1100 detected photons (a typical value in low signal-to-noise single molecule experiments) by steps of 50 nm  along the depth-of-focus. Signal distribution were adjusted by an integrated Gaussian fitting using readily available ImageJ plug-in ThunderSTORM \cite{ovesny_thunderstorm_2014}. A total of 50 frames were acquired at each step, to extract statistical repeatability (see Material and Methods). Note that to achieve diffraction-limited resolution, the pupil of the high NA objective had to be reduced (leading to an effective NA of $\sim 1,3$) to limit the collection of supercritical angle fluorescence arising from index mismatch at the sample interfaces and generating SA-like aberrations \cite{enderlein_imaging_2011}.

The first two panels of Figure \ref{locpres}a show the transverse profile of the PSF in the absence or presence of spherical aberrations. As expected, in the diffraction-limited (DL) case (aberration-corrected PSF), defocus occurs rapidly from the best plane of focus by widening the PSF (Figure \ref{locpres}b) which quickly reduces the localization precision of out-of-focus fluorescent emitters (Figure \ref{locpres}c). 

We now consider the case of the presence of spherical aberrations (SA) which arise from inhomogeneous convergence of nonparaxial rays after passing through a lens. For a converging lens, the marginal rays will be bent in front of paraxial rays, forming an ring-like envelope around a central profile called the caustic \cite{hecht_optics_2017}. Obviously, the caustic will affect super-localization performances along the depth-of-focus. In our experiments, we artificially introduced SA by mean of the correction collar of our objective.  As a result, we observe three regimes that are clearly highligted when considering the full-with-at-half-maximum (FWHM) of Gaussian fits performed along the recorded PSF (Figure \ref{locpres}b). 
Within the 1 \textmu m around the best focus plane (corresponding to the circle of least confusion), the PSF spread is the smallest. Along this axial region, we achieve close to diffraction-limited fits, and obtain localization precision around 10 nm (for 1100 detected photons) calculated from the standard deviation of the 50 measurement replicas at each plane.  This is a slightly lower performance compared to the DL-PSF but at the benefit of a modest DOF increase. At the axial tail of the PSF along the marginal focus i.e. for axial position lower than $\sim -$1.5 \textmu m, fitted PSF FWHMs split into two populations: a broad fitting artefact of the caustic ring having FWHM values greater than 700 nm and the fit of the narrow central distribution yielding FWHM values around 200 nm. Interestingly, by applying a threshold on the FWHM  by keeping only values below 700 nm, we can force the super-localization to the central profile. This allows to extend the DOF to this region even if the precisions are significantly lower (between 30 nm and 70 nm) compared to the ones obtained in the circle of least confusion ($\sim$ 10 nm) . In between these two regimes i.e. for z between $\sim -1.2$ and $\sim$-1.5 \textmu m, the caustic merges with the paraxial focus which causes a lateral broadening of the signal and leads to high FWHM values. In this area localization precision is around 40 nm which is four fold worse than at the circle of least confusion.

\begin{figure}
    \centering
    \includegraphics[scale=1]{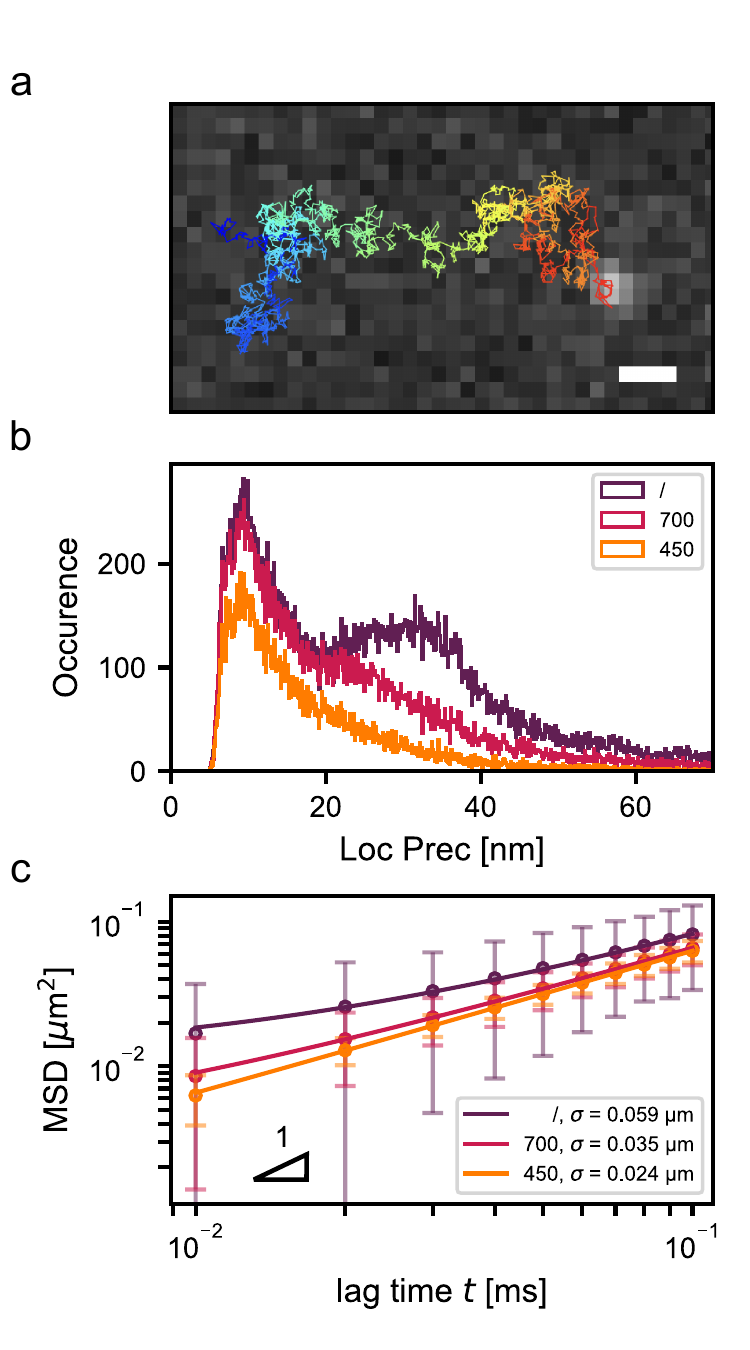}
    \caption{\textbf{SPT experiments in the presence of spherical aberrations} \textbf{(a)} Schematic of reconstructed trajectories. \textbf{(b)} Computed localization precision obtained from Gaussian fits over 15,000 frames without (\textbf{ / }) or with 700 nm and 450 nm FWHM thresholds. \textbf{(c)} Localization precision $\sigma$ of the reconstructed trajectories for the various thresholding values derived from the offsets of the average MSD plots.}
    \label{sa}
\end{figure}

To further confirm this analysis, we compared localization precision achieved by Gaussian fitting to the computed Cramer Rao Lower Bound (CRLB) \cite{ober_localization_2004} of the imaged PSFs (Figure \ref{locpres}c and Material and Methods). We observe that localization precision can be obtained at the CRLB limit at the best focus while in the caustic aberrant portions of the PSF, localizations obtained by 2D Gaussian fitting deviate from CRLB values indicating that in these regions, more complex fitting approaches should be applied in order to optimize localization precision. They could be based on e.g. maximum-likelihood estimation (MLE) with either a carefully tailored model function or with phase retrieved model PSF \cite{aristov_zola-3d_2018,ferdman_vipr_2020} but at the price of computational time and potential instabilities in the low-photon regime. Overall, in the presence of spherical aberrations, super-localization  can be obtained at high precision in the central portion of the SA-PSF while the tail region could in principle allow a slight extension of the DOF, bearing in mind that the PSF distribution varies along the axial direction.

To further inspect these properties, we next examined the localization performance of SA-PSF in single-particle tracking studies. Figure \ref{sa} presents analysis of SPT experiments performed on 100 nm fluorescent nanoparticles immersed in glycerol/water mixture (1100 photons are detected per particle in each frame using 10 ms integration time). Figure \ref{sa}a displays a reconstructed trajectory with a typical frame. In Figure \ref{sa}b, we present the localization precision computed from the fit parameters as described in the Materials and Methods section. For this we only retained fitted Gaussian having FWHMs below given thresholds as indicated in the figure. Indeed, with no FWHM thresholding, we observe that in this dynamic experiment, the caustic creates a binary distribution with values centered around 10 nm and 30 nm in accordance with Figure \ref{locpres}c. The distribution of localization precisions centered around 30 nm is broader than the one found around 10 nm, and comes from the inhomogeneous population of detections having FWHMs above 700 nm, as can be seen in Figure \ref{locpres}b. The effect of this broad distribution of localization values is also reflected in Figure \ref{sa}c where we present the computed means square displacements (MSDs) of the reconstructed trajectories (N = 198, 124 and 89 trajectories for the case of no, 700 and 450nm threshold on the FWHM respectively ; trajectory lengths range from 50 and 1435 data points). Indeed, without FWHM thresholding a large offset of the MSD, expressed by $4\sigma^2$, is observed at low time lags and is defined as the intercept of the MSD at t = 0, fitted by a linear function. 
Interestingly, this allows to recover the  $\sigma$ value of 53  nm which corresponds to the  localization precision of the emitters \cite{michalet_mean_2010}. It is reduced to 26 nm, and even 8 nm when a threshold of 700 nm and 450 nm respectively, is applied on fitted Gaussian FWHMs. The two latter values correspond to three and two times the diffraction limit of the microscope respectively.

From this observation, we conclude that in SPT experiments performed in the presence of spherical aberration, eDOF is subject to low localization precision due to strong variations of the PSF shape. Yet, we show in the following that the introduction of simple binary phase mask can regularize the PSF shape to obtain a wider eDOF without complex postprocessing steps.  

\begin{figure*}
    \centering
     \includegraphics[scale=0.9]{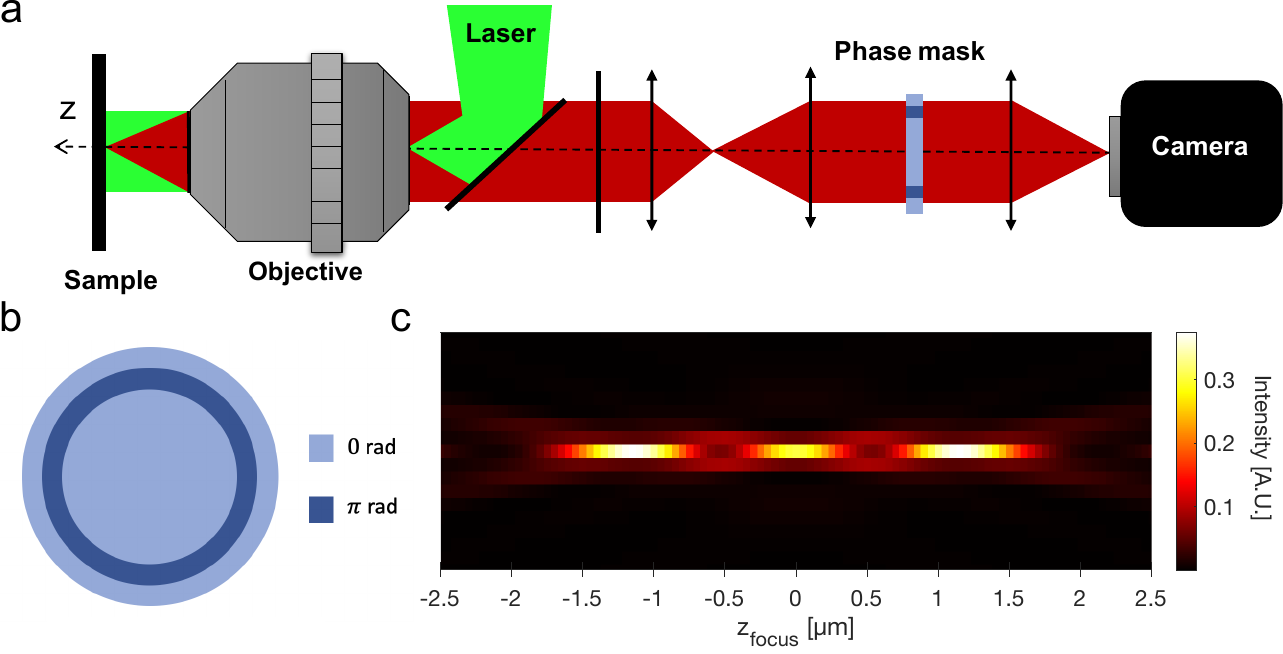}
    \caption{\textbf{Binary annular phase mask.} \textbf{(a)} Schematic of the widefield microscope where a phase mask is placed in a \textit{4-f} imaging relay and conjugated to the objective back focal plane. \textbf{(b)} Design of the binary annular phase mask used in this study. \textbf{(c)} Theoretical axial intensity profile of the eDOF-PSF.}
    \label{setup}
\end{figure*}

\subsection{Binary annular phase masks}

We now consider the presence of a phase mask in the pupil of the microscope to engineer the PSF of single emitters (Figure \ref{setup}a). The mask design is simple and consists of an annular binary phase-shift (Figure \ref{setup}b and Materials and Methods). This mask was specifically optimized for the 60x/1,45NA objective to allow super-localization on extended ranges \cite{leveque_co-designed_2020}.  In the absence of spherical aberration, we showed theoretically  that this mask generates over four micrometers eDOF but with varying localization precision values caused by a beaded structure of higher and lower intensities (Figure \ref{setup}c). More precisely, the simulated eDOF PSF displays  two types of regions: (i) some characterized by intense central lobes; (ii) others displaying a ring-like distribution with vanishing intensity in the center. It was further shown \cite{leveque_co-designed_2020} that the use of \textit{ad-hoc} position-dependent Maximum Likelihood Estimation would be required to attain comparable CRLB values in these two regions. In addition, at the transition between them, higher value of the CRLB would be achieved. Of note, this mask which creates beaded structures along the axis resembles other pupil engineering strategies such as Bessel droplet PSFs used to extend the DOF in point scanning microscopy by creating multiple foci in the excitation beam \cite{Chen_bessel_2022, he_tomographic-encoded_2022}.

\subsection{eDOF of single emitter localization using a binary phase mask}

We now investigate the performance of the annular binary phase mask for localizing single emitters in the presence of spherical aberrations. Interestingly, we observe that the presence of the aberrations becomes an asset as they regulate the PSF beaded structure along the axis (Figure \ref{locpres}a) and improves the overall performance of the system to achieve super-localization analysis by 2D Gaussian fitting at all axial position of the eDOF. 

To quantify this observation, we compared the results obtained with or without the phase mask.  More precisely, the fitted PSF FWHM displayed in Figure \ref{locpres}b becomes almost invariant, being narrow over more than four microns. This efficient regularization generates  PSF profiles that can adequately be fitted by simple Gaussian distributions yielding localization precision ranging from 12 nm to 27 nm along the considered range. Note that in this experiment,  the  insertion of the mask in the detection path slightly reduced  the photon budget to 700.

With this photon budget, we computed the CRLB which clearly points towards an efficient extension of the DOF but at the price of a slightly lower precision than obtained without the mask in the central region.  In addition, in eDOF areas where high signals are detected, super-localization is experimentally achieved at the CRLB limit whereas in the dimmer zones the measured localization precision deviates from the CRLB  by around 40$\%$. Note that this excellent result shows that some information contained in the PSF is not fully retrieved using Gaussian fitting in the regions where a ring PSF would occur without the presence of SA. Yet, spherical aberrations efficiently regularises the PSF generated by the annular binary phase mask.  Interestingly, this reasoning can also be  reversed by stating that the introduction of the annular binary phase mask allows to regularize and improve the eDOF PSF induced by spherical aberration. In any case, the use of annular binary phase mask in conjunction of spherical aberration constitutes a simple optical strategy to achieve eDOF superlocalization over more than four microns under low photon budget and without the need to tailor complex post-processing steps. In the following, we demonstrate some benefits of this approach in the context of SPT experiments. 

\begin{figure*}[htbp]
    \centering
    \includegraphics[scale=0.99]{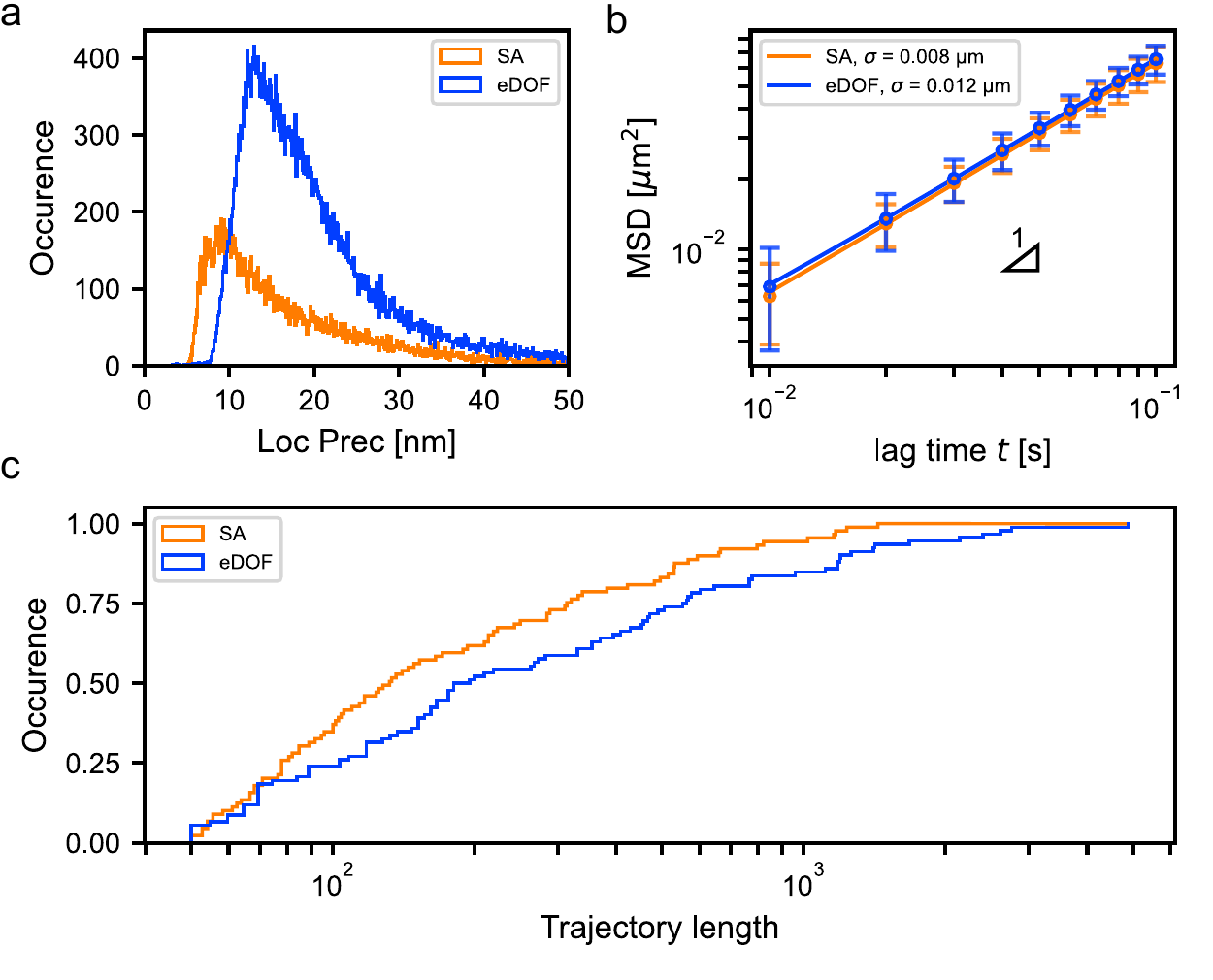}
    \caption{\textbf{eDOF Single-Particle Tracking} \textbf{(a)} Localization precision computed from Gaussian fits of fluorescent particles diffusing in liquid environment (15,000 frames) with (eDOF) and without (SA) using the binary annular phase mask. \textbf{(b)} Localization precision $\sigma$ of the reconstructed trajectories derived from the offsets of the average MSD plots. \textbf{(c)} Cumulative distribution of the lengths (number of frames) of the reconstructed trajectories. N= 46,085 localizations (resp 23,436) in 92 (resp. 89) trajectories in the presence (resp. absence) of the phase mask.}
    \label{trajlength}
\end{figure*}

\subsection{Extended trajectory length recordings in single-particle tracking experiments}

In common SPT experiments, emitters diffusing out of the imaging plane create discontinuities in the reconstructed trajectories or simply terminate them. These gaps which limit the length of the trajectories reduce the accuracy of particle diffusion analysis, including MSD and diffusion constant determination \cite{saxton_single-particle_1997}. We thus expected our simple approach to be effective in this context. For this, we tracked fluorescent nanoparticles imaged  in 2:1 Glycerol-Water mixture at 10 micrometers from the sample surface using $10\ ms$ integration time during 15,000 frames (see Materials and Methods). In these conditions,  we compared the performances of SA alone and eDOF in the presence of the annular binary mask. Trajectories were reconstructed following the application of a 450 nm threshold on the FWHM fitted Gaussian functions (see discussion above).

Figure \ref{trajlength}a displays the distribution of localization precision obtained from 46,085 (resp. 23,436) single-particle localizations in the presence (resp. absence) of the phase mask. Both distributions are narrow and almost unimodal. It is seen that the precision achieved with SA-PSF  (peaked at around 10 nm) is better than with eDOF-PSF (peaked at  around 15 nm) due to the lower photon budget induced by the presence of the mask and in accordance with Figure \ref{locpres}c. Note that the annular binary phase mask allows to double the number of detection events in this experiment at a same emitter concentration and acquisition duration. The average MSDs obtained from 92 (resp. 89) trajectories are similar yet with a small offset at short time lags consistent with the slightly lower localization precision in the presence of the phase mask (Figure \ref{trajlength}b).  More precisely,  analysis of the offsets (defined in section \ref{section_sa}) allows the dynamic localization error to be quantified and we obtain 8 nm localization precision with the SA-PSF and 12 nm in the case of the eDOF-PSF consistent with the analysis displayed in Figure \ref{locpres}a. In addition, the slope of the two MSDs provide comparable diffusion constant validating the adequacy of our approach to perform eDOF in the context of SPT. 

In addition, Figure \ref{trajlength}c demonstrates the increase in trajectory lengths obtained using  eDOF-PSF. More precisely, eDOF-PSF increases the median trajectory length by 44 $\%$, from 195 points to 395 points. More strikingly the use of the phase mask allows the collection of longer trajectories than can otherwise be obtained reaching up to 4,900 points and 10$\%$ of the trajectories longer than 1,200 points, which is a important benefit for SPT applications.

\section{Conclusion}
In this work, we have demonstrated that the introduction of a binary annular phase mask into a single-molecule fluorescence microscope can regularize  inhomogeneities in the PSF generated by spherical aberration and extend the DOF of super-localization of single emitters in a straightforwad manner. Indeed, the resulting PSF becomes restricted to a narrow distribution instead of exhibiting the typical caustic behavior of  spherical aberration.  This  allows applying  simple 2D Gaussian fitting procedure readily available  and  commonly used in localization-based super-resolution microscopy community. To demonstrate the effectiveness of this approach, we presented  SPT data recorded from nanoparticles diffusing in a liquid environment.  Trajectory lengths were increased by  capturing localization on eDOF. Future work will be required to determine the proper combination of spherical aberration amplitude and binary phase mask design needed to optimize specific eDOF ranges. We also anticipate that other sources of aberrations e.g. chromatic aberration might generate similar regularization of the PSF. In conclusion, we foresee that in the context of SPT experiments, the ability to significantly increase trajectory lengths may be of critical importance for many applications, as long trajectories contain the most information.

\section*{Materials and Methods}

\label{materials}

\subsection*{Optical setup}

The optical setup consists in conventional widefield microscope with a 60x/1,45NA Oil immersion objective (Nikon). We used 100 nm diameter fluorescent particles (TetraSpeck beads) adsorbed on a glass coverslip in PBS buffer and illuminated at 568 nm (Sapphire laser, Coherent) at $0.15\ kW/cm^2$. A 635 nm long pass fluorescence filter was placed at the exit of the microscope to select a fluorescence band of the TetraSpeak beads centered around 675 nm ($\sim$30 nm width). An imaging relay consisting of two achromat doublet lenses was placed after the microscope in order to insert the phase mask in the Fourier plane of the emitter image, thus conjugated to the back focal plane of the objective. The binary phase mask is placed on a kinematic mount with $\sim 1\mu m$ precision and aligned with respect to the microscope pupil with the insertion of a Bertrand lens. 
After passing the engineered pupil, the image is focused on an EMCCD camera (ProEM, RoperScientific) having 16 \textmu m pixels and using an EM Gain of 100. Following ADU calibration of the camera, the detected signals could be converted to photoelectron counts. 

\subsection*{Binary annular Phase Mask design and fabrication}

The binary annular phase mask is composed of 3 rings introducing alternatively a dephasing of  0 and $\pi$ at  the nominal wavelength $\lambda=700$ nm, as illustrated in fig 3.b. It has a clear aperture of radius 4.25 mm, and the outer radii of its first two rings have been optimized to reach the prescribed eDOF [10]. Their values are $\rho_1=2.906$ mm and $\rho_2=3.561$ mm. The mask  has been manufactured on quartz substrates using UV photolithography associated with Ion Beam Etching (IBE) followed by Inductive Coupled Plasma (RIE - ICP) etching. The etching depth $h$ is calculated so that the phase shift between two successive rings is equal to $\pi$, according to the formula:
\begin{equation}
\label{Step}
h = \frac{\lambda}{2(n-1)}
\end{equation}
where $n=1.4553$ is the refractive index of quartz at $\lambda = 700$ nm.
This leads to a required etching depth of $h = 769$ nm. 

\subsection*{PSF characterization and single emitter localization}
Z-stacks were acquired using 50 nm steps by computer control of the objective focus. At each step, 50 frames were acquired to generate statistical replicas with 10     ms exposure time per frame. Adjustment of the spherical aberrations was performed by the correction collar of the objective. Axial profiles displayed in figure \ref{locpres}a were constructed from the average of the 50 frames per step. In order to perform the super-localization, we use the ThunderSTORM Fiji plug-in where we set fitting parameters to default. After generation of the results, we kept only fitted PSF based on PSF FWHM thresholds as described in the main text. Localization precision was  computed as the standard deviation of the position of the super-localization retrieved from ThunderSTORM at each focus. 

We also computed the Cramér-Rao Lower bound from the Fisher Information Matrix for each of the PSF as defined by \cite{ober_localization_2004}:

\begin{equation}
\label{crlb}
\lbrack \textbf{I} (\theta) \rbrack_{ij} 
= \sum_{k=1}^{K}\frac{1}{b_k + \mu_\theta (k)}
\frac{ \partial \mu_\theta (k)}{\partial\theta_i}
\frac{\partial \mu_\theta (k)}{\partial\theta_j}
\end{equation}

where $K$ is the number of pixels, $\mu_\theta (k)$ is the is signal on the $k$th pixel, $\theta $ is the set of parameters and the noise is considered as Poissonian with mean $b_k + \mu_\theta (k)$ with a value retrieved from each experiments, with $b_k$ typically in the order of 7 photons.

\subsection*{Single-Particle Tracking}
Fluorescent particles (100 nm TetraSpeck beads) were diluted in a 2:1 Glycerol-Water mixture. Movies were acquired 10 \textmu m deep from the surface of the coverslip, over 15,000 frames at 10 ms exposure time. The illumination intensity was similar than for the PSF characterization ($0.15\ kW/cm^2$). Gaussian fitting was performed as described above using ThunderSTORM and thresholded accordingly (450 nm FWHM limit). Localizations were then imported into Python routines (based on TrackPy library \cite{trackpy}) to perform the linking step and generate single-particle trajectories. The search range and memory were set to 3 pixels and 3 frames respectively. Only trajectories above 50 frames were computed. Localization precisions present in trajectories were extracted from ThunderSTORM and are defined as proposed by \cite{quan_localization_2010}:

\begin{equation}
\label{eqtloc}
\left\langle(\Delta x)^{2}\right\rangle=\frac{2\sigma^{2}+a^{2}/12}{N}+\frac{8\pi\sigma^{4}b^{2}}{a^{2}N^{2}}    
\end{equation}

where $\sigma$ is the FHWM of the Gaussian fit, $a$ is the image pixel size, $N$ is the number of detected photons, and $b$ is the background noise.

\section*{DECLARATION OF COMPETING INTEREST}

The authors declare no competing financial interest.

\section*{ACKNOWLEDGEMENTS}
This work was performed with financial support from the European Research Council Synergy grant (951294), Agence Nationale de la Recherche (ANR-18-CE09-0019-02), the France-BioImaging National Infrastructure (ANR-10-INBS-04-01), the Idex Bordeaux (Grand Research Program GPR LIGHT) and the EUR Light S\&T (PIA3 Program, ANR-17-EURE- 0027) to L.C.. A.L. acknowledges support from the Fondation ARC pour la recherche sur le cancer. C.K and F.G. acknowledges support from CNRS through the research network (GdR ISIS) and the Exploratory Research Project MASK.







\end{document}